\documentstyle[11pt,amssymb,epsf]{article}

\makeatletter
\@addtoreset{equation}{section}
\makeatother

\def\ben{\begin{equation}}
\def\een{\end{equation}}

  \let\n=\nu  \let\p=\pi

\let\C=\Chi

\def\nn{\nonumber} \def\bd{\begin{document}} \def\ed{\end{document}}
\def\ds{\documentstyle} \let\fr=\frac \let\bl=\bigl \let\br=\bigr
\let\Br=\Bigr \let\Bl=\Bigl
\let\bm=\bibitem
\let\na=\nabla
\let\pa=\partial \let\ov=\overline
\newcommand{\be}{\begin{equation}}
\newcommand{\ee}{\end{equation}}
\def\ba{\begin{array}}
\def\ea{\end{array}}
\def\ft#1#2{{\textstyle{\frac{\scriptstyle #1}{\scriptstyle #2} } }}
\def\fft#1#2{{\frac{#1}{#2}}}
\def\del{\partial}
\def\vp{\varphi}
\def\sst#1{{\scriptscriptstyle #1}}
\def\oneone{\rlap 1\mkern4mu{\rm l}}
\def\td{\tilde}
\def\wtd{\widetilde}
\def\ie{{\it i.e.\ }}
\def\dalemb#1#2{{\vbox{\hrule height .#2pt
        \hbox{\vrule width.#2pt height#1pt \kern#1pt
                \vrule width.#2pt}
        \hrule height.#2pt}}}
\def\square{\mathord{\dalemb{6.8}{7}\hbox{\hskip1pt}}}
\newcommand{\ho}[1]{$\, ^{#1}$}
\newcommand{\hoch}[1]{$\, ^{#1}$}
\newcommand{\bea}{\begin{eqnarray}}
\newcommand{\eea}{\end{eqnarray}}
\newcommand{\ra}{\rightarrow}
\newcommand{\lra}{\longrightarrow}
\newcommand{\Lra}{\Leftrightarrow}
\newcommand{\bp}{\tilde \beta^\prime}
\newcommand{\tr}{{\rm tr} }
\newcommand{\Tr}{{\rm Tr} }
\def\0{{\sst{(0)}}}
\def\1{{\sst{(1)}}}
\def\2{{\sst{(2)}}}
\def\3{{\sst{(3)}}}
\def\4{{\sst{(4)}}}
\def\5{{\sst{(5)}}}
\def\6{{\sst{(6)}}}
\def\7{{\sst{(7)}}}
\def\8{{\sst{(8)}}}
\def\n{{\sst{(n)}}}
\def\cA{{{\cal A}}}
\def\cB{{{\cal B}}}
\def\cF{{{\cal F}}}
\def\cH{{{\cal H}}}
\def\tV{\widetilde V}
\def\tW{\widetilde W}
\def\tH{\widetilde H}
\def\tE{\widetilde E}
\def\tF{\widetilde F}
\def\tA{\widetilde A}
\def\im{{{\rm i}}}
\def\tY{{{\wtd Y}}}
\def\ep{{\epsilon}}
\def\vep{{\varepsilon}}
\def\bD{{{\bar D}}}
\def\R{{{\mathbb R}}}
\def\C{{{\mathbb C}}}
\def\H{{{\mathbb H}}}
\def\CP{{{\mathbb C}{\mathbb P}}}
\def\RP{{{\mathbb R}{\mathbb P}}}
\def\Z{{{\mathbb Z}}}
\def\bA{{{\mathbb A}}}
\def\bB{{{\mathbb B}}}
\def\bC{{{\mathbb C}}}
\def\bD{{{\mathbb D}}}
\def\bE{{{\mathbb E}}}
\def\bZ{{{\mathbb Z}}}
\def\Re{{{\frak{Re}}}}
\def\Im{{{\frak{Im}}}}
\def\cosec{{\,\hbox{cosec}\,}}
\def\Gm{{\Gamma_{\!\! -}}}
\def\Gp{{\Gamma_{\!\! +}}}
\def\stan{{standard }}
\def\nonstan{{supernumerary }}
\def\p{{\partial}}
\def\kdel#1{{\fft{\del}{\del#1}}}

\def\bog{{Bogomolny }}

\begin{document}

\begin{flushright}
MCTP-10-19\ \ \ \ MIFPA-10-23\\
May 2010
\end{flushright}

\vspace{10pt}
\begin{center}

{\Large {\bf Inconsistency of Breathing Mode Extensions of Maximal
Five-Dimensional Supergravity Embedding }}

\vspace{20pt}

James T. Liu$^{\dagger}$ and C.N. Pope$^{\ddagger,\star}$

\vspace{20pt}

{\hoch{\dagger}\it Michigan Center for Theoretical Physics\\
Randall Laboratory of Physics, The University of Michigan\\
Ann Arbor, MI 48109--1040, USA}

\vspace{10pt}

{\hoch{\ddagger}\it George P. \&  Cynthia W. Mitchell Institute
for Fundamental Physics and Astronomy\\
Texas A\&M University, College Station, TX 77843--4242, USA}

\vspace{10pt}

{\hoch{\star}\it 
 DAMTP, Centre for Mathematical Sciences,
 Cambridge University,\\  Wilberforce Road, Cambridge CB3 OWA, UK}

\vspace{40pt}

\underline{ABSTRACT}
\end{center}

  Recent work on consistent Kaluza-Klein reductions on Einstein-Sasaki
spaces prompted an intriguing conjecture that there might exist a
consistent $S^5$ reduction of type IIB supergravity to give 
five-dimensional ${\cal N}=8$ gauged supergravity coupled to a massive
supermultiplet that includes the breathing-mode scalar.  Motivated by this,
we investigate the possibility of augmenting the usual ${\cal N}=8$ 
supergravity reduction to include a breathing-mode scalar, and we
show that this is in fact inconsistent. The standard
reduction to the massless ${\cal N}=8$ supermultiplet depends for its
consistency on a delicate interplay between properties of the
ten-dimensional type IIB theory and properties of the Killing vectors
on $S^5$.  Our calculations show that turning on the breathing-mode
is sufficient to destroy the balance, and hence render the reduction
inconsistent.

\newpage

\section{Introduction}

    Kaluza-Klein reductions provide a mechanism for obtaining 
lower-dimensional theories from higher-dimensional ones.  The simplest
example is the reduction of a $(D+1)$-dimensional theory to $D$ 
dimensions, by taking the extra coordinate $z$ to lie on a circle.  If
one expands the $z$-dependence of all higher-dimensional fields in terms
of modes on the circle, then one is effectively just Fourier transforming
the original higher-dimensional theory.  However, in this example
one can additionally perform a {\it consistent truncation}, in which
all the infinite towers of massive fields, associated with non-trivial
$z$-dependent modes on the circle, are set to zero.  The crucial point
about this truncation is that setting the massive fields to zero is 
consistent with their own equations of motion.  As a result, the
remaining lower-dimensional theory, in which only the fields associated
with $z$-independent modes on the circle are retained, represents a 
consistent embedding within the original theory.  In other words,
any solution of the truncated lower-dimensional theory lifts back
as a solution in the higher dimension.

  There is a simple group-theoretic way of seeing the consistency of the
truncation described above.  The truncation retains all the singlets,
and only the singlets, under the action of the $U(1)$ isometry group of
the circle.  For example, a higher-dimensional field $\Phi(x^\mu,z)$
can be expanded as
\bea
\Phi(x^\mu,z) = \sum_{n=-\infty}^\infty \phi_n(x^\mu)\, e^{\im n z/L}\,,
\eea
where $L$ is the radius of the circle.  The singlet field, $\phi_0$,
is uncharged, while the $\phi_n$ fields with $n\ne 0$ occur in 
$\phi_{\pm n}$ charged doublets.  Clearly products of uncharged fields
can never act as sources for charged fields, and so setting all the
charged fields to zero will necessarily be a solution of their equations
of motion.

  The above argument can be generalised to the situation where a 
higher-dimensional theory is reduced on a manifold $M$ with isometry group
$G$.  If the full infinite towers of modes in a generalised Fourier
expansion are retained, then the resulting lower-dimensional theory is
necessarily a consistent embedding within the higher-dimensional one.  These
modes will carry representations of $G$. 
If one truncates the full system so as to keep all the singlets and
only the singlets under some subgroup $K\subset G$, then this truncation 
will be guaranteed to be a consistent one.  The reason is the same as in
the previous circle-reduction example; non-linear terms built from 
the $K$-singlets cannot act as sources for the truncated non-singlet
modes, and so setting all the non-singlets to zero is an exact solution
of their equations of motion.  

   A further point is that if the subgroup $K$ of the isometry group
acts transitively on $M$, then the number of $K$-singlets will be
finite.  This is generally what is wanted in a consistent truncation.
Typically, these singlet modes will include some (but perhaps not all)
of the massless fields in the full Kaluza-Klein towers, together, 
possibly, with some massive fields.

   A completely different kind of consistent truncation arises in
some special situations, typically, but not always, associated with
certain sphere reductions of supergravity theories.  The most well-known
example is the $S^7$ reduction of eleven-dimensional supergravity.
There exists an (almost cast-iron) proof that it is consistent to 
truncate the infinite towers of Kaluza-Klein modes so as to retain
precisely the fields of four-dimensional ${\cal N}=8$ $SO(8)$ gauged
supergravity \cite{de Wit:1986iy}.  There is no known group-theoretic
reason why this 
truncation is consistent; it depends for its success on very remarkable
and delicate 
conspiracies between properties of eleven-dimensional supergravity
and properties of the seven-sphere.  Other examples of these ``non-trivial''
consistent reductions include the $S^4$ reduction of eleven-dimensional
supergravity (for which there exists a complete proof of the consistency
\cite{Nastase:1999cb,Nastase:1999kf}),
and the $S^5$ reduction of type IIB supergravity (where there are strong
indications of the consistency
\cite{Khavaev:1998fb,Cvetic:1999xp,Lu:1999bw, cvluposatr}).

    One way of seeing how the properties of the supergravity theory and the
sphere conspire to make possible a consistent truncation was exhibited
for the $S^7$ reduction in \cite{dunipowa}.  The argument focuses on a
specific subset of the interactions in the 
full reduction procedure, namely those associated with the back reaction 
of the $SO(8)$ gauge fields on gravity in the lower dimension.
Prior to making any truncation, the four-dimensional fields will
include a tower of massive spin-2 fields, 
with the massless graviton at the bottom.  These fields arise from the 
mode expansion of the four-dimensional components of the eleven-dimensional
metric, in which the mode functions on the $S^7$ are the eigenfunctions
of the scalar Laplacian. {\it A priori}, one might expect that quadratic
products of the $SO(8)$ gauge bosons would contain not only a
singlet mode (\ie the usual Yang-Mills energy-momentum tensor) which would
act as a source for the massless graviton, but also a non-singlet 
contribution that would act as a source for massive spin-2 fields.  This
would prevent one from consistently setting the massive spin-2 fields to zero. 

   In fact, this is exactly what would happen if a {\it generic} theory
of gravity were reduced on a sphere.  The special features of 
eleven-dimensional supergravity and the seven-sphere that save the day 
are that the quadratic product of gauge bosons couples to spin-2
fields through two completely different sources, one being via the
Kaluza-Klein metric reduction ansatz, and the other via the reduction
ansatz for the 4-form field strength of eleven-dimensional supergravity.
When the two contributions are taken together, the non-singlet terms in
the quadratic product of gauge fields cancel out, and so the gauge bosons
excite only the (retained) massless graviton, but not the (truncated) massive
spin-2 fields.

   This is only a {\it necessary} condition for the full consistency of
the truncation to the massless sector, but it is a very revealing one.  
In fact the only known cases where reductions pass this test are precisely
those where in fact a fully-consistent reduction is either known to
exist, or strongly believed to exist.

   In this paper, we shall make use of an extension of the consistency
test in \cite{dunipowa} in order to study an intriguing recent conjecture that
even more remarkable consistent truncations might be possible 
\cite{Gauntlett:2009zw,gaunt}.  This
suggestion has arisen as a result of recent constructions of certain
consistent reductions of eleven-dimensional supergravity on Einstein-Sasaki
7-manifolds \cite{Gauntlett:2009zw}, and of type IIB supergravity on
Einstein-Sasaki 5-manifolds
\cite{Cassani:2010uw,Liu:2010sa,gaunt,Skenderis:2010vz}.
These reductions involve a consistent truncation to a set of
modes that includes massive as well as massless fields.  

   Taking the type IIB reductions as an example, the new consistent truncation
can be stated most straightforwardly in the case that the compactifying 
Einstein-Sasaki space $M_5$ is simply $S^5$.  The isometry group of $S^5$ is
$SO(6)$, but this contains an $SU(3)$ subgroup that acts transitively 
on $S^5$.  This is easily seen from the fact that $S^5$ can be viewed 
as a $U(1)$ bundle over $CP^2$, and this admits a natural action by
isometries of $SU(3)\times U(1)$, where $SU(3)$ acts on $CP^2$ and 
$U(1)$ acts on the fibres.  The statement of the new consistent reduction
is that one retains all the singlets, and only the singlets, under $SU(3)$.
The truncation is therefore guaranteed to be a consistent one.  Note that
included in this truncation are two massive scalar modes, associated with
an overall ``breathing'' deformation of the $S^5$, and a ``squashing''
of the $U(1)$ fibres relative to the $CP^2$ base \cite{Liu:2000gk}.

   Although the consistency of this reduction is obvious when the compactifying
space $M_5$ is just $S^5$, it is less obvious, though nevertheless true, that
$S^5$ can be replaced by {\it any} five-dimensional Einstein-Sasaki space,
with essentially the same reduction ansatz, and consistency is again 
achieved.  (One can describe any Einstein-Sasaki space as a $U(1)$ bundle
over an Einstein-K\"ahler base, and the modes analogous to those that 
are retained in the
$S^5$ reduction can be constructed in an almost identical fashion when
$CP^2$ is replaced by any other Einstein-K\"ahler base, since only
the invariant K\"ahler form and holomorphic 2-form play a r\^ole.)

   Motivated by these recently-constructed consistent reductions, it
has been suggested in \cite{Gauntlett:2009zw,gaunt} that it might be possible
to augment
the full ${\cal N}=8$ non-trivial consistent reduction of type IIB
supergravity on $S^5$ by including in addition certain massive 
supermultiplets.  In particular, one of these massive supermultiplets
would include the breathing-mode scalar.  There would also be associated
massive spin-2 fields in the supermultiplet.

   There are various reasons why one might be sceptical about this possibility.
First of all, it would violate a widely-held belief that it is not possible
to have a consistent coupling of (a finite number of) massive spin-2
fields to gravity.  (Kaluza-Klein theories, with the complete towers of massive
fields, do, of course, routinely manage to couple massive spin-2 to gravity.  
But in such examples, it is an {\it infinite} number of massive spin-2 fields,
and this appears to provide the loophole to the standard lore.)

   A related observation is that the massive spin-2 fields in the Kaluza-Klein
towers are associated with scalar harmonics on $S^5$.  The massive spin-2
fields contained in the supermultiplet that includes the breathing mode
lie in the {\bf 20} representation of $SO(6)$. {\it A priori}, 
one would expect that
there would be cubic terms in the complete lower dimensional Lagrangian, 
describing the
coupling of two of these massive spin-2 fields in the {\bf 20} representation
to a third spin-2 field, where this third field could be in any of the
$SO(6)$ representations for scalar harmonics on $S^5$ that could arise in
the symmetric product of the two {\bf 20} representations, namely the {\bf 1},
{\bf 20} and the {\bf 105}.  In the first two cases, these cubic interactions
would be associated with the {\bf 20} massive spin-2 fields giving 
energy-momentum tensor contributions sourcing the massless graviton and the
{\bf 20} of massive spin-2 fields themselves.  Both of these cubic 
interactions would be perfectly consistent with the conjecture that the
breathing-mode multiplet could be retained in a consistent truncation.  The
third of the cubic interactions listed above, however, would correspond
to the {\bf 20} massive spin-2 fields sourcing the yet-higher mass {\bf 105}
of spin-2 fields.   This would then be the start of an infinite sequence
of interactions that would be inconsistent with having set the {\bf 105}
of massive spin-2 fields, and those of yet higher mass, to zero.  
Since there is only one term in the 
ten-dimensional theory that contributes to this class of cubic interactions,
there is no possibility of any delicate cancellation between contributions
that might conspire to project out the {\bf 105} representation in the
harmonic expansion.  The situation for these cubic interactions for
massive spin-2 fields is therefore quite different from the situation
for the gauge bosons in the $S^5$ or $S^7$ supergravity reductions
described earlier, where the fact that they enter in the higher-dimensional
equations from two distinctly different sources (the metric and the 
antisymmetric tensor) provided the opportunity for a delicate cancellation
that avoided the otherwise expected excitation of massive spin-2 fields.

   A further reason for scepticism is that evidence coming from the 
recently-constructed reductions that include massive fields is actually 
not very persuasive.  As we noted above, in the case where the internal
space is $S^5$, the consistency of the existing ${\cal N}=2$ reduction 
with the breathing-mode multiplet is in fact guaranteed
by group-theoretic considerations, and so does not provide any non-trivial
test of 
the kind of conspiracies that would be necessary if a 
reduction to ${\cal N}=8$ gauged supergravity plus the breathing-mode
multiplet were to be consistent.

   In this paper, we shall show by considering a specific class of 
interactions within a hypothetical $S^5$ reduction to the massless
supergravity multiplet plus breathing-mode multiplet that conditions necessary
for the consistency of the truncation are not fulfilled.  Specifically,
we shall show that the inclusion of the breathing mode
is sufficient to destroy
the delicate conspiracies that allowed the truncation purely to the massless
${\cal N}=8$ sector to work.
Supplementing the massless sector and the
breathing mode by the rest of the associated massive supermultiplet would
not aid matters.  Thus, we conclude that it is not possible to obtain a
consistent truncation which includes the breathing mode as well as the
usual massless modes of ${\cal N}=8$ gauged $SO(6)$ supergravity. 

   Although our analysis in this paper focuses on the $S^5$ reduction
of type IIB supergravity, analogous arguments should hold in other cases
too, such as the $S^7$ reduction of eleven-dimensional supergravity.  Thus,
it is to be expected that turning on the breathing mode in the
reduction of eleven-dimensional supergravity 
to four-dimensional ${\cal N}=8$ gauged $SO(8)$
supergravity will similarly destroy the consistency of the reduction. 

\section{Breathing Mode Consistency Condition}

As demonstrated in \cite{dunipowa} for $D=11$ supergravity and 
\cite{tsikas,Hoxha:2000jf} for IIB supergravity, the manner in which the
Kaluza-Klein gauge fields back-react on the lower dimensional metric
provides a rather non-trivial consistency condition on the reduction.
The complete ansatz for the consistent $S^5$ reduction of the gravity plus
self-dual 5-form sector of type IIB supergravity, yielding
gravity coupled to $SO(6)$ Yang-Mills with 20 scalar fields, 
was obtained in \cite{cvluposatr}.

  Here, we shall extend the analysis of the 
consistency  condition on the IIB supergravity reduction, 
by  augmenting the discussion of
\cite{tsikas,Hoxha:2000jf} to include a breathing-mode scalar.
The relevant fields of IIB supergravity are the metric $\hat G_{MN}$ and the
self-dual five-form field strength $\hat H_5$.  The IIB equations of motion
for these fields are
\begin{eqnarray}
\hat R_{MN}&=&\fft1{96}\hat H_{MPQRS}\hat H_N{}^{PQRS},\nonumber\\
d\hat H_5&=&0,\qquad\hat H=\hat*H_5.
\label{eq:iibans}
\end{eqnarray}
(We use hats to denote 10-dimensional quantities.)
In order to demonstrate a necessary condition for consistency, we may
work at the linearized level for the gauge bosons but include the
breathing mode in a non-linear manner.  We thus take the ansatz
\begin{eqnarray}
d\hat s^2&=&e^{5\phi}e^\alpha e^\beta\eta_{\alpha\beta}+e^{-3\phi}
(e^a-K^{Ia}A^I)(e^b-K^{Jb}A^J)\delta_{ab},\nonumber\\
\hat H_5&=&\hat G_\5+\hat*\hat G_\5,\qquad
\hat G_\5=4ge^{(\alpha+20)\phi}\epsilon_5
-\fft1{2g}e^{(\gamma+4)\phi}{*F}^I\wedge dK^I.\label{ans1}
\end{eqnarray}
Here $K^{Ia}=K^{Im}e_m^a$ are the orthonormal components of the Killing
vectors on the internal space.  The ansatz (\ref{ans1}) is that used
in \cite{tsikas,Hoxha:2000jf}, except that we
have now introduced a breathing-mode scalar $\phi$.  The factors in the
metric ansatz are chosen to yield an Einstein frame reduction, and also
serve to normalize the breathing mode.  Once this normalization is fixed,
we have introduced constants $\alpha$ and $\gamma$ parameterising the
breathing-mode dependence in the five-form ansatz.  (The offsets are
chosen for latter convenience.)

We first consider the $\hat H_\5$ equation of motion.  At the linearized
level in gauge fields, this separates into $d\hat G_\5=0$ and
$d\hat*\hat G_\5=0$.  The former gives simply $d(e^{(\gamma+4)\phi}*F^I)=0$,
which is the linearization of the lower-dimensional Yang-Mills equation of
motion.  For $d\hat*\hat G_\5=0$ we first compute
\begin{equation}
\hat*\hat G_\5=4ge^{\alpha\phi}(\omega_\5-A^I\wedge*K^I)
+\fft1{2g}e^{\gamma\phi}F^I\wedge*dK+\cdots,\label{starG}
\end{equation}
where we have only kept terms linear in the gauge fields, and where $\omega_\5$
is the volume form on the internal manifold.  This allows us to write
\begin{eqnarray}
d\hat*\hat G_\5&=&4gde^{\alpha\phi}\wedge(\omega_\5-A^I\wedge*K^I)
+\fft1{2g}d(e^{\gamma\phi}F^I)\wedge*dK\nonumber\\
&&+4ge^{\alpha\phi}A^I\wedge d{*K}^I
+\fft1{2g}e^{\gamma\phi}F^I\wedge(d{*dK}^I-8g^2e^{(\alpha-\gamma)\phi}\,{*K}^I).
\end{eqnarray}
Using the Killing vector identities
\begin{equation}
d{*K}=0,\qquad d{*dK}-2\Lambda\, {*K}=0,
\end{equation}
which follow from the Killing equation $\nabla_mK_n^I+\nabla_nK_m^I=0$
and the Einstein condition $R_{mn}=\Lambda g_{mn}$, we see that
\begin{equation}
d\hat*\hat G_\5=4gde^{\alpha\phi}\wedge(\omega_\5-A^I\wedge*K^I)
+\fft1{2g}de^{\gamma\phi}\wedge F^I\wedge*dK\nonumber\\
+4ge^{\gamma\phi}(1-e^{(\alpha-\gamma)\phi})F^I\wedge*K^I,
\label{eq:dhg}
\end{equation}
where we have made use of the linearised 
Bianchi identity $dF^I=0$ and where we have
taken $\Lambda=4g^2$.

Demanding that $d\hat*\hat G_\5=0$ requires the independent vanishing of
terms with different tensor structure in (\ref{eq:dhg}).  Starting with
$d\phi\wedge\omega_\5$, we see that $\alpha=0$.  After this, it is easy
to see that vanishing of the remaining terms demands $\gamma=0$.
Thus we conclude that $\hat*\hat G_\5$ cannot
have any dependence on the breathing mode $\phi$ at all.  Furthermore,
this conclusion
would be unchanged even if we had allowed for more complicated functional
dependence on $\phi$%
\footnote{At first sight, it might seem that our
conclusion that $\alpha$ must vanish was an artefact of our having neglected
to include any term in the ansatz for $\hat G_\5$ that involved $d\phi$, 
which might then have been able to cancel terms such as $d\phi\wedge\omega_\5$
arising in $d{\hat*}\hat G_\5=0$.  However, such an addition to the 
ansatz is not possible.  In order to cancel the $d\phi\wedge\omega_\5$
term in (\ref{eq:dhg}), it would be necessary to include a term of
the form $4\alpha g e^{\alpha\phi}\,d\phi\wedge
 \omega_\4$ in (\ref{starG}), where
$d\omega_\4=\omega_\5+\cdots$.  However, the required 4-form $\omega_\4$
on $S^5$ is not globally defined, and thus such a term cannot appear in
the ansatz.  See section 3 for an extended discussion of this point.}.

Having examined the equation of motion for $\hat G_\5$, we may now turn to
the lower-dimensional components of the IIB Einstein equation.  This in
particular allows us to examine the stress tensor associated with the
Yang-Mills fields $F^I$.  The reduction of the ten-dimensional Ricci tensor
corresponding to (\ref{eq:iibans}) yields
\begin{eqnarray}
\hat R_{\alpha\beta}&=&e^{-5\phi}(R_{\alpha\beta}-\ft52\eta_{\alpha\beta}
\Box\phi-30\partial_\alpha\phi\partial_\beta\phi
-\ft12e^{-8\phi}K^{Ia}K^J_aF^I_{\alpha\gamma}F^{J\,\gamma}_\beta),\nonumber\\
\hat R_{ab}&=&e^{-5\phi}(e^{8\phi}R_{ab}+\ft32\delta_{ab}\Box\phi
+\ft14e^{-8\phi}K^I_aK^J_bF^I_{\alpha\beta}F^{J\,\alpha\beta}),\nonumber\\
\hat R_{\alpha b}&=&-\ft12e^{-9\phi}K^I_b(D_\beta F^{I\,\beta}_\alpha
-8F^{I\,\beta}_\alpha\partial_\beta\phi),
\end{eqnarray}
while
\begin{equation}
\hat H_{\alpha NPQR}\hat H_\beta{}^{NPQR}=
-384g^2e^{15\phi}\eta_{\alpha\beta}+\fft{24}{g^2}
e^{-\phi}\nabla_aK^I_b\nabla^aK^{I\,b}(F^I_{\alpha\gamma}F^{J\,\gamma}_\beta
-\ft14\eta_{\alpha\beta}F^I_{\gamma\delta}F^{J\,\gamma\delta}).
\end{equation}
Putting everything together yields the lower-dimensional Einstein equation
\begin{eqnarray}
R_{\alpha\beta}-\ft12\eta_{\alpha\beta}R
&=&2g^2(5e^{8\phi}-2e^{20\phi})\eta_{\alpha\beta}
+30(\partial_\alpha\phi\partial_\beta\phi-\ft12\eta_{\alpha\beta}
\partial_\gamma\phi\partial^\gamma\phi)\nonumber\\
&&+\ft12e^{-8\phi}(F^I_{\alpha\gamma}F^{J\,\gamma}_\beta
-\ft14\eta_{\alpha\beta}F^I_{\gamma\delta}F^{J\,\gamma\delta})Y^{IJ},
\end{eqnarray}
where
\begin{equation}
Y^{IJ}=K^{I\,a}K^J_a+\fft1{2g^2}e^{12\phi}\nabla_aK^I_b\nabla^aK^{J\,b}
\label{eq:ydef}
\end{equation}
is constructed out of the 15 Killing vectors on the internal manifold.  The
first term in $Y^{IJ}$ comes from the metric reduction, while the second
arises through the five-form ansatz.  Note that, at this order,
we have ignored the coupling of the Yang-Mills fields to the non-breathing
mode scalars.  Although the full Einstein equation will involve these scalars,
demonstrating the inconsistency of turning on a breathing mode will not
require knowledge of these scalar couplings.

Consistency of the Kaluza-Klein truncation demands that $Y^{IJ}$ be
independent of the internal coordinates $y^m$.  Even in the absence of
a breathing mode, this is a highly non-trivial condition, as it involves
the cancellation of $y^m$ dependence through the interplay of two terms.
Specifically, for the $SO(6)$ Killing vectors of $S^5$ we have
that $K^{Ia}\, K_a^J$ is a $y$-dependent 
$15\times 15$ matrix of rank 5, while $(\nabla_a K_b^I)(\nabla^a K^{Jb})$
is a $y$-dependent $15\times 15$ matrix of rank 10, but in their sum, with
precisely the $(2 g^2)^{-1}$ coefficient in the second term, 
the $y^m$ dependence cancels and the result is a
constant matrix, of rank 15. 
By choosing the basis and normalisation for
the Killing vectors appropriately, this constant rank-15 matrix can
be chosen to be proportional to $\delta^{IJ}$.

If the breathing mode $\phi$ is now turned on, it is evident that the
previous consistency that required the precise balancing of the 
two Killing vector expressions in (\ref{eq:ydef}) will now fail.
One way to express this is that consistency then demands a condition of the
form
\begin{equation}
Y^{IJ}=\beta(\phi)\delta^{IJ},
\label{eq:ycond}
\end{equation}
where $\beta(\phi)$ is a function of $\phi$ alone, and not of the internal
$y^m$ coordinates.  Given the consistency of the truncation in the absence of
a breathing mode, so that $Y^{IJ}=\beta(0)\delta^{IJ}$ holds when $\phi=0$,
we may rewrite the above condition as
\begin{equation}
(1-e^{12\phi})K^{I\,a}K^J_a=
\left(\beta(\phi)-e^{12\phi}\beta(0)\right)\delta^{IJ}.
\end{equation}
This condition is clearly violated in general whenever the breathing mode
is active, as the left hand side depends on $y^m$, while the right hand side
does not.  We have
thus shown that it is inconsistent to perform a breathing-mode reduction
of IIB supergravity on $S^5$ that entails the coupling of the ${\mathcal N}=8$
gauged $SO(6)$ supergravity to a breathing-mode supermultiplet.

Note that the consistent breathing-mode Sasaki-Einstein reductions 
\cite{Gauntlett:2009zw,Cassani:2010uw,Liu:2010sa,gaunt,Skenderis:2010vz}
involve a single Abelian graviphoton corresponding to the gauging of the
preferred $U(1)$ isometry of the Sasaki-Einstein space.  In the absence
of a breathing mode, the consistency of such reductions involving $U(1)$
bundled over a K\"ahler-Einstein base was studied in \cite{Hoxha:2000jf}.
A convenient manipulation of (\ref{eq:ydef}) yields
\begin{equation}
Y^{IJ}=(1+2e^{12\phi})K^{I\,a}K^J_a+\fft1{4g^2}e^{12\phi}\Box(K^{I\,a}
K^J_a).
\end{equation}
The $U(1)$ graviphoton is associated with a Killing vector of constant
length on $S^5$, which corresponds to taking $K^{I\,a}=\delta^{a9}$
(assuming 9 to be the $U(1)$ fiber direction), in which case the second
term above vanishes.  The result is $Y=1+2e^{12\phi}$, which yields a
consistent truncation with a lower dimensional Lagrangian of the form
\cite{Cassani:2010uw,Liu:2010sa,gaunt,Skenderis:2010vz}
\begin{equation}
{\mathcal L}_5=R*1-\ft12(1+2e^{12\phi})F_2\wedge*F_2+\cdots.
\end{equation}
Finally, while we are primarily focused on supersymmetric breathing
mode reductions, the condition (\ref{eq:ycond}) is universal, regardless
of supersymmetry.  We have thus shown that turning on a breathing mode
is incompatible with retaining the non-Abelian gauge bosons corresponding
to the gauging of a non-trivial isometry of the internal space.  This
result is of course compatible with the known breathing-mode reductions
that either throw out the non-Abelian gauge bosons through a non-supersymmetric
truncation \cite{brdulupost}, or work in an ${\mathcal N}=2$ context with
only a single Abelian graviphoton
\cite{Gauntlett:2009zw,Cassani:2010uw,Liu:2010sa,gaunt,Skenderis:2010vz}.

\section{An Explicit Example}

   In the previous section, we gave a general discussion that demonstrated
the inconsistency of including the breathing mode in the $S^5$ reduction
of type IIB supergravity with the $SO(6)$ gauge bosons. In this section, 
we shall consider a simplified sector within this analysis, in which only
a $U(1)$ subgroup of the $SO(6)$ gauge fields is active.  This will allow
us to present completely explicit reduction ans\"atze, and to elaborate 
on some of the issues encountered in the previous section.  We should
emphasise that the $U(1)$ subgroup that we shall consider here
is {\it not} the one associated with the graviphoton considered in
\cite{Gauntlett:2009zw,Cassani:2010uw,Liu:2010sa,gaunt,Skenderis:2010vz},
and our discussion in this section will show that the inconsistency
of including the breathing mode in an $S^5$ reduction can be exhibited
even with just a single suitably chosen $U(1)$ gauge field turned on.

   We take as our starting point the consistent $S^5$ 
reduction of type IIB supergravity that yields five-dimensional
${\cal N}=4$ gauged $SU(2)\times U(1)$ supergravity, which was studied in
\cite{lupotr}.  This is a consistent truncation of the full ${\cal N}=8$
gauged $SO(6)$ reduction.

  The five-dimensional bosonic fields of the consistent reduction exhibited
in \cite{lupotr} comprised the metric; the gauge fields of $SU(2)\times U(1)$;
a single scalar field $X$; and a complex 2-form potential.  For the present
purposes, we may consistently set the $SU(2)$ gauge fields and the
complex 2-form potential to zero.  The only ten-dimensional fields
that are non-vanishing for this simplified reduction ansatz are
the metric $\hat g_{MN}$ and the self-dual 5-form $\hat H_\5$, satisfying
the equations of motion (\ref{eq:iibans}).
 From \cite{lupotr}, the consistent reduction ansatz is then given by\footnote{
Note that the Killing vector associated with the $U(1)$ gauge field 
we are considering here is $\del/\del\tau$, which has length $X\Delta^{-1/4}
\,g^{-1}\, s^{-1}$, and so it is not constant on $S^5$ (even if $X=1$).  This 
contrasts with the $U(1)$ graviphoton 
gauge field retained in the consistent reduction
considered in 
\cite{Gauntlett:2009zw,Cassani:2010uw,Liu:2010sa,gaunt,Skenderis:2010vz}, 
which is associated with the Killing vector describing translations along
the $U(1)$ fibers and which does have constant length.}  
\bea
d\hat s_{10}^2 &=& \Delta^{1/2}\, ds_5^2 + g^{-2}\,\Big( X\, \Delta^{1/2}\,
d\xi^2 +  X^2\, \Delta^{-1/2}\, s^2\, (d\tau - g\, A_\1)^2 \nn\\
&&\qquad\qquad\qquad\quad +
   \Delta^{-1/2}\, X^{-1}\, c^2\, d\Omega_3^2\Big)\,,\label{metans1}\\
\hat G_\5 &=& 2 g U\, \ep_\5 - \fft{3sc}{g}\, X^{-1}\, {*dX}\wedge d\xi
- \fft{sc}{g^2} X^4\, {*F_\2}\wedge d\xi\wedge\omega\,,\label{G5ans1}
\eea
where
\bea
\Delta&=& X^{-2}\, s^2 + X\, c^2\,,\qquad 
U= X^2\, c^2+ X^{-1}\, s^2 + X^{-1}\,,\qquad
s\equiv \sin\xi\,,\qquad c\equiv\cos\xi\,,\nn\\
\omega&=& d\tau - g\, A_\1 \,,\qquad F_\2=dA_\1\,,
\eea
and the self-dual 5-form is written as $\hat H_\5=\hat G_\5 + {\hat *}
\hat G_\5$.  Note that $d\Omega_3^2$ denotes the metric on the unit
3-sphere. It is useful also to record that the ten-dimensional
dual of $\hat G_\5$ is
\be
\hat *\hat G_\5= -\fft{2s c^3}{g^4}\, U \Delta^{-2}\,  
 d\xi\wedge\omega\wedge\Omega_3 + 
 \fft{3 s^2 c^4}{g^4}\, X^{-2}\, \Delta^{-2}\, dX\wedge\omega\Omega_3
-\fft{c^4}{g^3}\, X\Delta^{-1}\, F_\2\wedge \Omega_\3\,,\label{G5dual1}
\ee
where $\Omega_\3$ is the volume form of the unit 3-sphere metric
$d\Omega_3^2$.

   Substituting the ansatz into the ten-dimensional equations of
motion, one finds that the resulting five-dimensional equations can be
derived from the Lagrangian \cite{lupotr}
\be
{\cal L}_5 = R {*\oneone} - 3 X^{-2}\, {*dX}\wedge dX - \ft12 
  X^4\, {*F_\2}\wedge F_\2+ 4g^2\, (X^2+ 2 X^{-2})\, {*\oneone}\,.
\ee
Note from (\ref{G5ans1}) and (\ref{G5dual1}) that the equation of 
motion for the self-dual 5-form, $d\hat H=0$, has the independent 
consequences that $d\hat G_\5=0$ and $d{\hat *}\hat G_\5=0$.

  We now investigate the effect of adding the 
breathing-mode scalar $\phi$.  In the metric
ansatz, this will modify (\ref{metans1}) as in (\ref{ans1}) to give
\bea
d\hat s_{10}^2 &=& e^{5\phi}\, \Delta^{1/2}\, ds_5^2 + 
g^{-2}\, e^{-3\phi}\, \Big( X\, \Delta^{1/2}\,
d\xi^2 +  X^2\, \Delta^{-1/2}\, s^2\, (d\tau - g\, A_\1)^2 \nn\\
&&\qquad\qquad\qquad\qquad\qquad\qquad +
   \Delta^{-1/2}\, X^{-1}\, c^2\, d\Omega_3^2\Big)\,.\label{metans2}
\eea
In the ansatz for the 5-form, it is easiest first to consider
$\hat *\hat G_\5$, for which we write
\be
\hat *\hat G_\5= -\fft{2s c^3}{g^4}\, e^{\alpha\phi}\, U \Delta^{-2}\,
 d\xi\wedge\omega\wedge\Omega_3 +
 \fft{3 s^2 c^4}{g^4}\, e^{\beta\phi}\, (X 
\Delta)^{-2}\, dX\wedge\omega\wedge \Omega_3
-\fft{c^4}{g^3}\, e^{\gamma\phi}\, X\Delta^{-1}\, F_\2\wedge \Omega_\3\,,
\label{G5dual2}
\ee
where $\alpha$, $\beta$ and $\gamma$ are constants to be determined.  The
ansatz for $\hat G_\5$ itself can now easily be calculated, and we find
\bea
\hat G_\5 &=& 2 g U\, e^{(\alpha+20)\phi}\, \ep_\5 - 
\fft{3sc}{g}\, e^{(\beta+12)\phi}\, X^{-1}\, {*dX}\wedge d\xi
- \fft{sc}{g^2} \, e^{(\gamma+4)\phi}\, 
X^4\, {*F_\2}\wedge d\xi\wedge\omega\,,\nn\\
&&\label{G5ans2}
\eea
(We postpone until later in this section the discussion of the possible
addition of terms involving $d\phi$ in the ansatz.)

   As in the case without the breathing mode, the ten-dimensional equation
$d\hat H_\5=0$ leads to the separate equations $d\hat G_\5=0$ and
$d{\hat *}\hat G_\5=0$.  Looking at the various independent 6-forms in
$d{\hat *}\hat G_\5=0$, we find from the coefficients of
$dX\wedge d\xi\wedge\omega\wedge\Omega_3$, $dX\wedge F_\2\wedge\Omega_3$, 
$d\xi\wedge F_\2\wedge\Omega_3$ and $d\phi\wedge d\xi\wedge\omega\wedge\Omega_3$
respectively that $\alpha=\beta$, $\beta=\gamma$, $\alpha=\gamma$, and
$\alpha=0$. Thus we conclude that $\alpha=\beta=\gamma=0$.

  Our interest is to look for an unambiguous 
signal of inconsistency (or consistency) in
the putative reduction process in the presence of the breathing-mode scalar.
To this end, we shall focus on a specific class of terms coming from
the reduction procedure, namely those contributions of the form
\be
R_{\alpha\beta} \sim F_{\alpha\gamma}\, F_{\beta}{}^\gamma +\cdots
\ee
in the lower-dimensional Einstein equation.  These terms quadratic
in the $U(1)$ gauge field come from two sources; one being from
the ansatz (\ref{metans2}) for the reduction of the ten-dimensional metric,
and the other being from the $F_\2$ terms in the ansatz for $\hat H_\5$.
In \cite{lupotr}, the ten-dimensional Ricci tensor is calculated for
the reduction ansatz (\ref{metans1}) without the breathing mode.  After
straightforward calculations, we find that for the metric reduction
ansatz (\ref{metans2}) that includes the breathing mode, the 
contribution to the lower-dimensional components of the higher-dimensional
Ricci tensor of the form $F_{\alpha\gamma}\, F_{\beta}{}^\gamma$ is
\be
\hat R_{\alpha\beta} = \Delta^{-1/2}\, e^{-5\phi}\, R_{\alpha\beta}
 - \ft12 s^2\, X^2\, \Delta^{-3/2}\, e^{-13\phi}\, F_{\alpha\gamma}
  F_\beta{}^\gamma +\cdots\,.
\ee

    From the ansatz (\ref{G5dual2}) and (\ref{G5ans2}) (with, as we
then learned, $\alpha=\beta=\gamma=0$), we find the contribution
\be
\hat H_{\alpha BCDE}\,\hat H_\beta{}^{BCDE} =
48 c^2 e^{-\phi}\, X^5\, \Delta^{-3/2}\, F_{\alpha\gamma} F_{\beta}{}^\gamma
  + \cdots\,.\label{Hsquared}
\ee
Thus, from the ten-dimensional equations of motion (\ref{eq:iibans}) we
find
\be
R_{\alpha\beta} = \ft12 X^4\, \Delta^{-1}\, e^{4\phi}\, 
( X^{-2} \,s^2 e^{-12\phi} + X c^2)\, F_{\alpha\gamma} F_\beta{}^\gamma
  + \cdots\,.\label{d5einst}
\ee
It should be emphasised that we have focused just on this specific
type of structure, namely terms proportional to 
$F_{\alpha\gamma} F_\beta{}^\gamma$, and {\it no other terms that we 
have omitted could alter the form of these contributions.}  It is clear,
therefore, that we have run into an inconsistency, since the left-hand side
of (\ref{d5einst}) depends only on the lower-dimensional coordinates,
whilst the right-hand side depends also on the coordinate $\xi$ on the
internal five-sphere.  (Recall that $s=\sin\xi$ and $c=\cos\xi$.) 
This is a classic example of what can go wrong in 
an inconsistent truncation.

   It should be noted that if we had not introduced the breathing
mode (\ie if we had considered just the reduction investigated in
\cite{lupotr}), then all would have been well.  With $\phi=0$, the
$\xi$ dependence of the two terms on the right-hand side would cancel,
since then the factor in parentheses is just $(X^{-2}\, s^2 + X c^2)$,
which equals $\Delta$, thus cancelling the ($\xi$-dependent) $\Delta^{-1}$
factor.

  We now return to the question of whether terms involving $d\phi$ 
could have been introduced in the ansatz for $\hat G_\5$, thereby 
circumventing the conclusion arrived at in the discussion
below (\ref{G5ans2}), namely that $\alpha=\beta=\gamma=0$.  
After some experimentation, we find that if we add the term
\be
{\hat *}\hat G_\5^{\rm extra} = \fft{\alpha c^4}{g^4}\, e^{\alpha\phi}\,
  X \Delta^{-1}\, d\phi\wedge\omega\wedge\Omega_3\label{G5dualextra}
\ee
to (\ref{G5dual2}), then we satisfy $d{\hat *}\hat G_\5=0$ identically,
provided only that we impose $\alpha=\beta=\gamma$.  In fact, we then have
that with this extra term included,
\be
{\hat*} \hat G_\5 = d\Big( \fft{c^4}{g^4}\, e^{\alpha\phi}\,
  X\Delta^{-1}\, \omega\wedge\Omega_\3\Big)\,.
\ee

  With the apparent freedom now to allow $\alpha\ne0$, which would introduce
an extra factor of $e^{2\alpha\phi}$ in (\ref{Hsquared}) and in the
second term in (\ref{d5einst}), it might
seem that we could now restore consistency in the Einstein equations
by choosing $\alpha=-6$ so that the the $\xi$-dependence on the
right-hand side of (\ref{d5einst}) cancels, yielding
\be
R_{\alpha\beta} = \ft12 X^4 e^{-8\phi}\, F_{\alpha\gamma} F_\beta{}^\gamma
   +\cdots\,.
\ee

   However, the extra term (\ref{G5dualextra}) that needed to be added is
actually singular, as may be seen by calculating the associated vielbein
components, which are 
\be
 ({\hat *}\hat G_\5^{\rm extra})^{\phantom \chi}_{\alpha 6123}=
\fft{\alpha c}{s}\, \Delta^{-5/4}\, e^{(\alpha+7/2)\phi}\,
  X^{-7/4}\, \del_\alpha\phi\,,
\ee
(where the 1, 2 and 3 directions are on $S^3$, and 6 is along the
$\omega$ direction).  Thus the vielbein components blow up at the north
and south poles of the 5-sphere, where $\sin\xi=0$.  This is a concrete
realisation of the point alluded to in section 2, that the needed
extra term in the ansatz that might have allowed $\alpha$ to be
non-zero is proportional to $d\phi\wedge \omega_\4$, where $\omega_\4$
is a (necessarily singular) 4-form whose exterior derivative yields
the volume form of $S^5$.

   There are in fact even worse problems associated with 
the introduction of the extra term (\ref{G5dualextra}). To see these, it
suffices to set $X=1$ and $A_\1=0$, and focus on the contributions
of the breathing-mode scalar in the lower-dimensional components of the
ten-dimensional Einstein equations.  We find that these are given by
\be
R_{\alpha\beta} = \ft58 \Box\phi\, g_{\alpha\beta} + 30\del_\alpha\phi\,
\del_\beta\phi + \fft{\alpha^2 c^2}{4 s^2}\, e^{(2\alpha+12)\phi}\,
(2\del_\alpha\phi\, \del_\beta\phi - (\del\phi)^2\, g_{\alpha\beta})\,,
\label{einst3}
\ee
where the terms on the right-hand side that are proportional to $\alpha^2$
are the
contribution from (\ref{G5dualextra}). 
The $\xi$-dependence of these terms
implies that we have a classic example of an inconsistent reduction.
This problem would not, of course, be remedied by restoring the
$X$ and $A_\1$ fields that we omitted in this simplified discussion.

  The upshot of this discussion is that the conclusion reached in the
discussion following eqn (\ref{G5ans2}) is robust; the constants
$\alpha$, $\beta$ and $\gamma$ should all be zero, and hence from 
eqn (\ref{d5einst}) we have indeed demonstrated the inconsistency of
retaining the breathing mode as well as the $U(1)$ gauge field of this
truncation.

\section{Conclusions}

   In this paper we have investigated the idea, motivated by
the conjecture made in
\cite{gaunt}, that it might be possible to augment the
usual consistent $S^5$ reduction of type IIB supergravity to
five-dimensional massless
${\cal N}=8$ gauged $SO(6)$ supergravity, by including as well the
massive supermultiplet associated with the breathing-mode scalar.  Our
conclusion is that such an extended reduction is, unfortunately, 
not a consistent one.  In fact our result is somewhat stronger, and
quite independent of supersymmetry.  Namely, we have shown that the
consistency of the reduction to the massless bosonic sector (including the
gauge fields of $SO(6)$) is destroyed as soon as one tries to augment the
ansatz to include the breathing-mode scalar.

   An essential ingredient in our argument is that one can focus on
certain subsets of interactions within the
full reduction procedure, which would not in
themselves provide fully complete and consistent reductions, provided one
is careful to extract only those conclusions that are insensitive to
the presence or absence of the omitted terms.  In this spirit, we extended
a consistency test that was first developed long ago to probe the 
validity of including the non-Abelian gauge bosons in the $S^7$ reduction
of $D=11$ supergravity and the $S^5$ reduction of type IIB supergravity.
The consistency of these reductions depends on highly non-trivial
conspiracies involving the details of the higher-dimensional theories and
the properties of the Killing vectors on the compactifying spheres.
In our extension of this analysis, we showed in the case of the $S^5$ 
reduction of type IIB supergravity, these delicate conspiracies are 
destroyed if the breathing mode is introduced as well.  

   The conclusion
is that neither the breathing mode, nor its entire associated
massive supermultiplet, can be included in any enlargement of the 
existing consistent reduction of type IIB supergravity to five-dimensional
${\cal N}=8$ gauged $SO(6)$ supergravity.  Analogous arguments would
lead to a similar conclusion for the $S^7$ reduction of eleven-dimensional
supergravity.

\section*{Acknowledgments}

We are grateful to Mike Duff and Jerome Gauntlett for useful discussions.
JTL wishes to acknowledge the hospitality of the Mitchell Institute for
Fundamental Physics and Astronomy where this work was initiated, and the
hospitality of the LPTHE Jussieu and IPhT CEA/Saclay where this work was
completed.
This work was supported in part by the US Department of Energy under grants
DE-FG02-95ER40899 and DE-FG03-95ER40917.


\end{document}